\documentclass[twocolumn,10pt, final]{asme2ej}

\usepackage{epsfig} 
\usepackage{cite}
\usepackage{amssymb}
\usepackage{amsfonts}
\usepackage{amsmath}
\usepackage{mathrsfs}
\usepackage{enumitem}
\usepackage[ruled, linesnumbered, vlined]{algorithm2e}
\usepackage{caption}
\usepackage{subcaption}

\setlist[description]{font=\normalfont\itshape}

\newcommand\blfootnote[1]{%
	\begingroup
	\renewcommand\thefootnote{}\footnote{#1}%
	\addtocounter{footnote}{-1}%
	\endgroup
}

\DeclareSymbolFont{newfont}{OML}{cmm}{m}{it}
\DeclareMathSymbol{\Epsilon}{3}{newfont}{15}

\makeatletter
\def\step{%
	\@ifnextchar[ \@step{\@noitemargtrue\@step[\@itemlabel]}}
\def\@step[#1]{\item[#1]\mbox{}\\}
\makeatother

\title{Fast One-to-Many Multicriteria Shortest Path Search}

\author{Temirlan Kurbanov
	\affiliation{
		Artificial Intelligence Center\\
		Faculty of Electrical Engineering\\
		Czech Technical University in Prague\\
		email: kurbatem@fel.cvut.cz
	}	
}

\author{Marek Cuch\'{y}
	\affiliation{
		Artificial Intelligence Center\\
		Faculty of Electrical Engineering\\
		Czech Technical University in Prague\\
		email: cuchy@fel.cvut.cz
	}	
}

\author{Ji\v{r}\'{i} Vok\v{r}\'{i}nek
	\affiliation{
		Artificial Intelligence Center\\
		Faculty of Electrical Engineering\\
		Czech Technical University in Prague\\
		email: jiri.vokrinek@fel.cvut.cz
	}	
}

\begin{document}

\maketitle

\begin{abstract}
	{\it This paper introduces a novel algorithm combination designed for fast one-to-many multicriteria shortest path search. A preprocessing algorithm excludes irrelevant vertices by building a smaller cover graph. A modified version of multicriteria label-setting algorithm operates on the cover graph and employs a dimensionality reduction technique for swifter domination checks. While the method itself maintains solution optimality, it is able to additionally incorporate existing heuristics for further speedups. The proposed algorithm has been tested on multiple criteria combinations of varying correlation. The results show the introduced approach provides a speedup of at least 6 times on simple criteria combinations and up to 60 times on hard instances compared to vanilla multicriteria label-setting. Graph preprocessing also decreases memory requirements of queries by up to 13 times.
	}
\end{abstract}
	
	\blfootnote{This work was supported by the Research Center for Informatics, Czech
		Technical University in Prague, 16627 Prague, Czech Republic.}
	
	\blfootnote{This work has been submitted to the IEEE for possible publication. Copyright may be transferred without notice, after which this version may no longer be accessible.}
	
\section{Introduction}
Shortest path problem is one of the most well-known and studied tasks in graph theory. It poses a challenge of finding a feasible path between a pair of vertices that has a minimum possible sum of its component edge weights. However, the traditional form of the problem is only concerned with minimization of a single criteria parameter, which limits its applicability to real-world tasks. A prominent example of insufficient characterization provided by a single criterion is route planning in transportation. Edges in road networks can be assigned multiple parameters, such as traversal time, distance, fuel requirement, and each of the parameters can potentially influence optimality of a path. 

Therefore, multicriteria or multiobjective shortest path search emerged as a natural extension of the aforementioned task. In contrast to the single-criteria problem, multicriteria shortest path search is aimed at producing a path minimizing multiple objective parameters at the same time. Instead of a single path with minimum cost, its solution is represented by a Pareto set of optimal (pairwise non-dominated) paths. Multiple algorithms solving the shortest path problem have been successfully adapted to the multicriteria version (i.e. Dijkstra's algorithm \cite{Martins}, A* algorithm \cite{MOA*}). However, the size of a Pareto set can grow exponentially in the size of a graph \cite{complexity}, thus making the problem NP-hard. Moreover, the solution size is also exponential in the number of criteria considered. As a result, vanilla versions of the algorithms tend to have unacceptably high runtimes on larger graphs such as country-size road maps. 

To alleviate this issue, a number of acceleration techniques have been developed. Unfortunately, a major part of the techniques in consideration either can only be applied to one-to-one search (i.e. guiding heuristics for A*) or does not preserve the solution optimality. On the other hand, most research papers address mainly the bicriteria setting. And although the overwhelming majority of the methods can be applied to an arbitrary number of criteria, an increase in the parameter dimensionality poses further limitations.

The objective of this paper is to introduce a practical algorithmic approach tailored for accelerated one-to-many path search in the context of multicriteria route planning. The proposed method combines a swift preprocessing phase with dimensionality reduction \cite{dimensionality} aimed at dominance check simplification. As a result, it is capable of extracting optimal Pareto sets at least 6-8 times faster than a vanilla path search algorithm while having a reduced memory footprint.

The performance of the algorithm has been tested on multiple graphs structurally representing the road networks of Bavaria and San Francisco bay. Some of the objective parameters for the datasets have been generated artificially in order to test the algorithm on parameter combinations that vary not only in size, but also in correlation. This approach provided the possibility to analyze single parts of the combined algorithm and estimate their effectiveness in different possible scenarios.

The layout of this article is as follows. Section 2 provides a brief overview of the research that has been conducted on the topic and the related types of problems. Section 3 addresses the formal statement of one-to-many multicriteria shortest path problem. The theoretical description of the proposed algorithm along with the pseudocodes is laid out in section 4. For the transparency and ease of understanding, this section is divided into 4 subsections. The first three subsections address the individual parts of the combination, while the last one explains the way these are connected into a single algorithm. Section 5 is dedicated to the evaluation of the algorithm's performance on test datasets, and multiple aspects have been evaluated: preprocessing efficiency, runtime, and memory requirements. A conclusion to the article is provided in section 6.

\section{Related work}
It was stated earlier that multicriteria optimal path search traces back to the single shortest path search problem. In the same manner, the earliest algorithms solving the multicriteria version are based on Dijkstra's algorithm \cite{Dijkstra}. One of the first algorithms to solve the multicriteria task has been proposed by Martins \cite{Martins} in 1984. The multicriteria label-setting algorithm proposed by Martins maintains a Pareto set of optimal labels, expanding the lexicographically smallest one during every iteration. Lexicographic ordering allows the algorithm to only expand non-dominated labels, and once a label is expanded and closed, it is guaranteed to belong to the optimal Pareto set of a node. In contrast, the label-correcting approach \cite{Vincke} expands labels in the first-in-first-out (FIFO) fashion, compensating for the expansion of non-optimal labels with the time saved on ordering absence.

Goal-directed one-to-one techniques have also been extended successfully to manage multiple objective parameters. Namely, MOA* \cite{MOA*} is a modification of A* algorithm \cite{A*} designed specifically for multicriteria shortest path search. Its improved version called NAMOA* \cite{NAMOA*} has been proposed later.

The increasing importance of shortest path search in the domain of transportation gave rise to a multitude of algorithms and acceleration techniques designed to exploit the inherent features of road networks such as their quasiplanar structure. A comprehensive survey of routing algorithms and their application in road networks has been conducted by Bast \textit{et al.} \cite{Bast}. Among other techniques, it also considers some of the "state-of-the-art" algorithms such as Contraction Hierarchies (CH) \cite{CH} and Customizable Route Planning (CRP) \cite{CRP}. In its core, CRP is able to incorporate multiple criteria, including turn costs. On the other hand, while CH was originally designed to only consider a single criterion, there have been attempts to extend it to multicriteria cases \cite{MCH}.

Accelerating techniques for multicriteria shortest path search can be separated into two categories: optimality preserving or not. There are few optimality preserving techniques for the given problem, and fewer of them can be applied to one-to-many search. One of the existing methods is based on the observation that label-setting queries are parallelizable \cite{parallel}. The label-setting property of the algorithm remains if a subset of globally Pareto optimal labels are expanded in parallel. During every iteration, the approach identifies a subset of such labels and scans these in parallel. While this approach can be effectively implemented for bicriteria cases, its generalization to higher dimensions is non-trivial since no efficient way to identify and maintain globally optimal labels has been proposed yet.

The majority of existing acceleration techniques for multicriteria shortest path search are heuristics returning nearly optimal solutions. Some approaches are based on heuristic optimization methods, for example stochastic evolutionary algorithms \cite{stoch_ev, bi_ev}. Other ones employ standard label-setting or label-correcting techniques combined with relaxation heuristics providing substantial speedups at the cost of solution optimality. These include $\Epsilon$-dominance, cost-based pruning, ellipse pruning, etc. \cite{practical_routing} and operate by discarding solutions that are considered irrelevant based on their similarity or geographical features.

\section{Multicriteria shortest path search problem}
Multicriteria optimization is concerned with optimization of multiple objective criteria simultaneously. These objectives are often conflicting, which excludes the possibility of finding a single optimal solution. Instead, one can build a set of Pareto optimal solutions. A solution is called Pareto optimal if it cannot be improved in any of the criteria without degrading at least one of the other criteria. In other words, each of the Pareto optimal solutions provides a possible trade-off between the criteria. A Pareto set is thus a set of all Pareto optimal solutions. To formalize this notion, multicriteria optimization defines the dominance property. The relation of \textit{weak dominance} of a tuple $y'=\langle y_1', ..., y_q'\rangle \in \mathbb{R}^q$ by a tuple $y = \langle y_1, ..., y_q \rangle \in \mathbb{R}^q$ is defined as follows:
\begin{equation}
	\begin{gathered}
		y \succeq y' \textrm{ iff:} \\
		\forall i \in \{ 1, ..., q \}
		\begin{cases}
			y_i \leq y_i' & \textrm{ if } i \textrm{ is minimized}\\
			y_i \geq y_i' & \textrm{ if } i \textrm{ is maximized}.
		\end{cases}
	\end{gathered}
\end{equation}
Accordingly, $y$ \textit{dominates} $y'$ ($y \succ y'$) iff $y \succeq y'$ and $y \npreceq y'$.

Let $G = (V, E, I)$ be a finite labeled directed graph with $|V|$ vertices and $|E|$ edges. Every edge $e = (v_i, v_j) \in E$ starting at a vertex $v_i \in V$ and ending at $v_j \in V$ has an associated vector of criteria-costs $c_e \in \mathbb{R}^q$. A path in $G$ is any sequence of nodes $\pi = v_1, v_2, ..., v_k$ such that for all $i < k$, $(v_i, v_{i+1}) \in E$. A path can also be represented by its compound edges $\pi = e_1, e_2, ..., e_{k-1}$, where it holds $\forall i \in \{1, ..., k-1\}: e_i = (v_i, v_{i+1}) \in E$. The cost vector of $\pi$ is defined by a function $c(\pi)$ incorporating cost vectors of edges included in $\pi$. The most commonly used function of a path is the sum of the cost vectors of its component edges. However, this is not always the case, since the function is defined by the specific setting and the criteria in consideration. For example, for some path criteria such as surface quality, speed limit, and number of lanes it is more appropriate to maintain the average value instead of the sum. For a given source vertex $s$ and a subset $I \subseteq V$ of goal points, the task is to find the Pareto sets of optimal routes to all the goal vertices. The article addresses the "one-to-many" version of the problem as opposed to "one-to-all" due to the fact that the preprocessing phase removes the majority of vertices from the graph in order to provide speed and memory optimization. This assumption is sensible in real-world scenarios. For example, a driver planning a route to the nearest fuel station or pharmacy is not interested in routes to books stores, which eliminates the necessity to calculate and maintain Pareto sets for their corresponding vertices.

Although the notion of "multiple" objective criteria implies any amount of criteria from two and higher, the majority of the experiments dedicated to the topic are mainly conducted on graphs containing only two criteria. This subcategory of the problem in question is often called "bicriteria shortest path search". In research, most algorithms for multicriteria shortest path search problem can be tested on bicriteria problem instances without loss of generality, while decreasing the potential time and power requirements.

\section{Multicriteria shortest path search algorithm}
As it was stated earlier, the overall approach is a combination of three techniques: a graph preprocessing method, a query algorithm, and a dimensionality reduction technique exploited in queries. This section describes each of these independently and provides a layout of the combined approach afterwards.

\subsection{Graph preprocessing}
In modern path planning algorithms, graph preprocessing is a commonly used technique. Some preprocessing techniques are based on pruning and route precalculation, allowing query algorithms to ignore the parts of graphs that are not expected to provide value during the planning stage (\textit{e.g.} Geometric Containers \cite{GC}, ArcFlags \cite{ArcFlags}). Other preprocessing methods are aimed at graph compression, decreasing the memory requirements and runtimes of the search algorithm (\textit{e.g.} Vertex Separators \cite{VS}, CRP \cite{CRP}). An important benefit of preprocessing techniques is they should only be performed once on a single graph instance, which is especially useful when operating on large instances such as road networks. As long as the cost function and the structure of the graph are unchanged, its preprocessed version remains actual. This gives developers the possibility to perform a preprocessing operation with high resource demands in advance and save its results for subsequent usage.

\subsubsection{Vertex cover construction}
The approach in discussion uses a compressive preprocessing technique called \textit{k-(All-)Path Covers} (kPC) \cite{kPC}. For a given directed graph $G = (V, E,I)$ and a size constant $k \in \mathbb{N}_+$, a k-Path cover is a subset of vertices $C \subseteq V$ such that for every (not necessarily shortest) simple path $\pi = v_1, ..., v_k$ in $G$ it holds $C \cap \pi \neq \emptyset$. For a given graph and a size constant, there is a k-Path cover of minimum size. However, the complexity of finding the minimum k-Path cover has been proven to be APX-hard \cite{kPC_complexity}, which makes its generation intractable in practice. Thus, k-Path covers of suboptimal sizes are more appropriate for the method to be of use in real-world problem solving.

Since the naive enumerative approach for k-Path cover generation would have unacceptably high runtimes on larger graph instances, Funke \textit{et al.} \cite{kPC} proposed another approach based on iterative pruning. Initially, all the nodes in the graph are assumed to belong to the cover, i.e. $C = V$. The nodes are examined one by one in a chosen order. For every node $v \in C$, the algorithm tests if it must stay in the cover to maintain the path property. In order for the property to hold after the removal of $v$, there must exist no path of size $k$ containing $v$ such that it does not include any other vertex from $C$. If such a path is found, then $v$ remains in the cover. Essentially, to test this condition one must explore all incoming and outgoing paths of $v$ reaching other nodes of $C$. If there is a simple combination of an incoming and an outgoing path with cumulative size at least $k$, $v$ is kept in $C$.

A pseudocode of this method outlined in Algorithm \ref{kPCpseudocode} shows two stages. The algorithm is permitted from removing the subset of goal points $I$ from the cover (lines 3-4). During the first stage (lines 5-7), all simple paths outgoing from $v$ and ending at another cover node are enumerated. This can be performed using a two-step technique based on depth-first search (DFS). First, DFS starts from the examined vertex and traverses only outgoing edges without visiting other cover vertices. If a path with size $k$ has been found this way, the second step can be omitted, since the necessity of $v$ in $C$ has already been proven. Otherwise, the second step (lines 8-15) attempts to extend every enumerated path to $k$ by connecting it to an incoming path. This step is also based on the DFS method, though the edges must be traversed backwards. It is important to note that the algorithm only explores simple paths of length at most $k$, since $C$ is assumed to be a feasible cover before examination of $v$. The correctness of the algorithm can be proved trivially through mathematical induction.

\begin{algorithm}
	\caption{kPC construction}
	\label{kPCpseudocode}
	\KwData{graph $G=(V,E,I)$, path size $k$}
	\KwResult{kPC graph $C=(V',E')$}
	\BlankLine
	\tcc{node cover construction}
	$V' = V$ \\
	\For{$\forall v \in V'$}{
		\If{$v \in I$}{
			\textbf{continue}
		}  	
		$P_o$ = the set of all outgoing paths from $v$ not containing other nodes from $V' - \{v\}$ \\
		\If{$\exists \pi_o \in P_o$ such that $|\pi_o| \geq k$}{
			\textbf{continue}
		}
		remove  = \textbf{true} \\
		\For{$\forall \pi_o \in P_o$}{
			$\pi_i$ = an incoming path to $v$ not containing nodes from $(V' \cup \pi_o) - \{v\}$ such that $|\pi_i| + |\pi_o| - 1 = k$ \\
			\If{$\exists \pi_i$}{
				remove = \textbf{false} \\
				\textbf{break}
			}
		}
		\If{remove = \textbf{true}}{
			remove $v$ from $V'$
		}
	}
	\BlankLine
	\tcc{edge cover construction}
	\For{$\forall v \in V'$}{
		perform DFS search starting from $v'$ \\
		\If{a node $w \in V'$ is met during search}{
			save the current path to new edge $e$\\
			\For{every other cover edge $e'$ from $v$ to $w$}{
				\If{$e$ dominates $e'$}{
					remove $e'$ from $E'$\\
				}
				\ElseIf{$e'$ dominates $e$}{
					discard $e$\\
					\textbf{break}
				}
			}
			if $e$ is not discarded, add it to $E'$
		}
	}
\end{algorithm}

\subsubsection{Edge cover construction}
In order for a shortest path search algorithm to be able to operate on the cover, the cover vertices must be connected by overlay edges (lines 16-26). This is achieved in a fashion similar to cover extraction. For every vertex $v \in C$ in the completed cover, DFS search is started from it. Whenever the search meets another cover vertex $w \in C$, a new edge from $v$ to $w$ is added to the overlay graph with its cost vector being the cost function value of the found path. Further path expansion through $w$ is unnecessary. As a result, the overlay edges only connect pairs of directly adjacent cover vertices, i.e. cover vertices connected by a path in the original graph which does not contain any other cover vertex.

As one can observe, the resulting overlay graph is not simple due to there being multiple possible paths between a pair of non-adjacent vertices. Moreover, a significant part of the routes produced by this method is redundant in context of shortest path search, since DFS enumerates and saves all the possible paths, even the unnecessarily long and winding ones. Therefore, multiple edge pruning techniques have been described in \cite{kPC_pruning}, the essential one being domination pruning. In short, for a set of overlay edges $S$ between a pair of cover vertices, an edge $e' \in S$ is pruned from $S$ if there is another edge $e \in S$ such that $e \succeq e'$, i.e. $e$ (weakly) dominates $e'$. Standard domination pruning tests domination of an edge by every other edge in $S$, which gives $|S|^2$ domination checks for a pair of vertices. Since quadratic runtime would be very costly on large graphs with high $k$ size values, another approach can be used. For every considered criterion $i \in \{1, ..., q\}$, one can find the edge $e_i' \in S$ having the optimal value $c_i$ among all edges in $S$. Afterwards, all other edges in $S$ can be checked against these optimal edges, which would only yield $q|S|$ comparisons. Although dominance pruning can be performed after overlay edge construction is finished, a far more preferable option is to perform it online, whenever a new cover edge is formed (lines 20-25). Since the vast majority of the original cover edges produced by DFS are exceedingly long, these may have unacceptably high memory requirements. On the other hand, while online dominance pruning may require a higher number of operations, it allows the construction algorithm to operate on a much smaller amount of memory.

The second pruning technique taken from \cite{kPC_pruning} and used in experiments is triangle pruning. Let there be three cover vertices $u, v, w$ such that there are cover edges between the pairs $(u, w)$, $(u, v)$, $(v, w)$. Triangle pruning removes from the edge set of $(u, w)$ edges that are dominated by at least one combined path $(u, v, w)$. This pruning method can also be accelerated similarly to domination pruning. Out of all of the produced edge combinations of $(u, v)$ and $(v, w)$, one can extract the combinations with the optimal criteria values and evaluate the $(u, w)$ set against these. Unlike dominance pruning, which can be performed on the fly, triangle pruning must be performed after all of the cover edges are formed due to the fact that it operates on multiple cover edge sets at a time.

\subsection{Multicriteria label-setting algorithm}
Multicriteria label-setting algorithm (MLS) \cite{Martins} is one of the first introduced algorithms to solve the multicriteria shortest paths search problem. Despite that, it is still one of the most frequently used algorithms for one-to-all search. As it was stated earlier, it can be considered an extension of Dijkstra's algorithm to multicriteria cases due to their similar structure and operation.

However, MLS has several crucial differences. Firstly, instead of operating on vertex-distance pairs, MLS operates on \textit{labels}. A label $l = (v, (c_1(v), ..., c_q(v)))$ of a node $v$ contains ID of the vertex it belongs to and the criterion-cost vector $(c_1(v), ..., c_q(v))$ indicating the cost of its respective path from source node to $v$. Secondly, instead of keeping track of the minimum distance for every vertex in the graph, MLS maintains two sets of non-dominated labels associated to every vertex $v \in V$: a set of permanent labels $perm(v)$ and a set of temporary labels $temp(v)$. Permanent labels of a vertex remain unchanged and belong to its Pareto set, while temporary labels may be removed during the execution of the algorithm. Lastly, in order to expand at each iteration the most promising label, the algorithm maintains a priority queue of all temporary labels which are sorted in lexicographic order of their cost vectors. Given two vectors $c = (c_1, ..., c_q)$ and $c' = (c_1', ..., c_q')$, the notion of $c$ preceding $c'$ in lexicographic ordering is defined as
\begin{equation}
	\begin{gathered}
		c <_\ell c' \textrm{ iff:} \\
		\\
		\exists k \in \{1, ..., q\}:
		\{\forall i \in \{1, ..., k\}: \; c_i = c_i'\} \textrm{ and} \\
		\begin{cases}
			c_k < c_k' & \textrm{if } k \textrm{ is minimized}\\
			c_k > c_k' & \textrm{if } k \textrm{ is maximized}\\
		\end{cases}
		\\
		\textrm{OR}
		\\
		\forall i \in \{1, ..., q-1\}: c_i = c_i' \; \textrm{ and} \\
		\begin{cases}
			c_q \leq c_q' & \textrm{if } q \textrm{ is minimized}\\
			c_q \geq c_q' & \textrm{if } q \textrm{ is maximized}\\
			
		\end{cases}
	\end{gathered}
\end{equation}

MLS starts by generating a blank label for the source node and adding it to an empty lexicographically ordered priority queue and the temporary label set of the source node. Temporary and permanent label sets of all other vertices are empty. After that, the main algorithm loop starts to operate. At each iteration, the lexicographically smallest label $l = (v, (c_1(v), ..., c_q(v)))$ is extracted from the priority queue. It is also moved from the temporary set of its corresponding vertex to the permanent one. Label expansion proceeds as follows: for every adjacent vertex $w, (v, w) \in E$, a new label $l' = (w, (c_1(w), ..., c_q(w)))$ is created. It is then compared to the labels in $perm(w)$ and $temp(v)$. If $l'$ is dominated by at least one of these labels, it is discarded. On the other hand, if $l'$ dominates a label $t \in temp(w)$, $t$ is removed from $temp(w)$ and from the priority queue. It is worth mentioning that $l'$ cannot dominate any label from $perm(w)$. If the dominance tests are completed and $l'$ is not dominated, it is added to $temp(w)$ and to the priority queue. The algorithm terminates when the priority queue is emptied, and permanent sets of vertices contain their exact Pareto sets. 

In order for the algorithm to be able to reproduce the path corresponding to an individual entry of a Pareto set, every label must also include a pointer to its parent label, which is omitted in this description. The desired path can then be extracted by traversing labels through the parent pointers from the preferred entry of the goal vertex up to the source. The operation closely resembles Dijkstra's path extraction procedure. The pseudocode of the algorithm is presented in \cite{Martins}.

\subsection{Dimensionality reduction}
A significant runtime part of MLS as well as other multicriteria shortest path search algorithms is spent on dominance checks. These time requirements grow with the number of criteria to be considered. Although most multicriteria methods can be limited to bicriteria cases without loss of generality, in practice addition of a single criterion may have a dramatic effect on the overall performance of an algorithm. For this reason, a dimensionality reduction technique applicable to multicriteria label-setting algorithms has been developed \cite{dim, dimensionality}.

Given a vector $v = (v_1, ..., v_n)$, its \textit{truncated vector} $t(v)$ is $v$ without the first component, i.e. $t(v) = (v_2, ..., v_n)$. For a set of vectors $S$, its corresponding \textit{set of truncated vectors} is the set of non-dominated truncated vectors $T(S) = \{\forall t(v) \; | \; \nexists v': t(v') \succ t(v); \; v, v' \in V \}$. Dimensionally reduced weak dominance check is based on \textit{t-discard} operation: given a set of vectors $S$ and a vector $v$, $v$ is t-discarded by $S$ if for $\forall v' \in S: v'_1 \leq v_1$ and there is a vector $w \in S$ such that $t(w) \succeq t(v)$.

It has been proven in \cite{dimensionality} that t-discarding procedure can partially replace regular dominance checks in a multicriteria label-setting algorithm without losing solution optimality. The assumption is that at every iteration, the label selected for expansion is the lexicographically smallest among all the temporary labels available. This property ensures that when a label $l$ of a node $n$ is selected for expansion, all of its permanent labels stored in $perm(n)$ are lexicographically smaller than $l$. Validity of this assumption is ensured by extracting labels for expansion from a priority queue sorted in lexicographic order. Given this, regular dominance tests of $l$ against $perm(n)$ can be replaced by checking if $perm(n)$ t-discards $l$. In the case of weak dominance, the first condition holds automatically. If all the vectors in $perm(n)$ are lexicographically smaller than the vector of $l$, the maximum value of the first criterion in $perm(v)$ is necessarily not greater than that of $l$. On the other hand, when replacing regular dominance by t-discarding, it is preferable to maintain $perm(v)$ lexicographically sorted. This way, the validity of the first condition for t-discarding can be ensured by comparing the first criterion of $l$ to that of the last label in $perm(v)$. To check if the second condition holds, it is necessary to check the truncated criteria vector of $l$ against all vectors in $T(perm(n))$. However, this requires significantly less operations, since the dimension of vectors is decreased by one, and hence the size of $T(perm(n))$ is in most cases smaller than that of $perm(n)$ due to the non-dominance property of truncated sets.

The principle of t-discarding procedure causes a notable distinction between the bicriteria instances of multicriteria shortest path search problem and the instances with higher numbers of goal criteria. The truncated version of a vector consisting of two elements is a single number. Hence, for a collection of bicriterial vectors, the corresponding set of non-dominated truncated vectors is bound to only consist of a single element represented by a number. Thus, a check if a newly generated label is t-discarded by the set of permanent labels during a query can be performed via only 2 comparisons regardless of the set size. This property is not influenced by the meaning and correlation of the parameters in question. The observation implies an assumption of dimensionality reduction generally operating with higher efficiency and stability in bicriteria shortest path search scenarios.

\subsection{Combined algorithm}

After all key parts have been described, it is now possible to lay out the overall approach. Given a graph $G=(V, E, I)$ and a subset of interest points $I \subseteq V$, the algorithm first builds a kPC of the graph and interconnects it with non-dominated overlay edges. As it was stated earlier, the overlay graph $G' = (V', E')$ remains valid until the initial graph is changed, so the overlay can be read from memory if it was built beforehand. When the preprocessing phase is finished, the algorithm can perform queries on the prepared overlay. Algorithm \ref{tKpcMls} presents the query pseudocode.

\begin{algorithm}
	\caption{t-discarding kPC MLS search}
	\label{tKpcMls}
	\KwData{graph $G=(V,E,I)$, its kPC cover $C=(V',E')$, source vertex $s$}
	\KwResult{Pareto sets of paths from $s$ to the goal set $I$}
	\BlankLine
	\For{$\forall v \in V'$}{
		$perm(v) = \emptyset$\\
		$temp(v) = \emptyset$\\
		$tset(v) = \emptyset$\\
	}
	\textit{connectSourceToKpc(G, C, s)}\\
	\textit{queue} = empty lexicographically ordered priority queue\\
	\textit{queue.push(}source label\textit{)}\\
	
	\BlankLine
	\While{queue is not empty}{
		$l$ = \textit{queue.pop()}\\
		$v$ = $l$.node\\
		$temp(v).remove(l)$\\
		$perm(v).add(l)$\\
		$tset(v).update(l)$\\
		\BlankLine
		\For{$\forall w \in V'$ adjacent to $v$}{
			\For{$\forall e \in E'$ connecting $(v, w)$}{
				$l'$ = $addSuffix(l, e)$\\
				\If{path of l' is infeasible}{
					\textbf{continue}
				}
				\tcc{domination check}
				\If{tset(w) t-discards l'}{
					\textbf{continue}
				}
				isDominated = \textbf{false}\\
				\For{$\forall c \in temp(w)$}{
					\If{$c \succeq l'$}{
						isDominated = \textbf{true}\\
						\textbf{break}
					}
					\ElseIf{$l' \succeq c'$}{
						remove $c$ from $temp(w)$ and from \textit{queue}
					}
				}
				\If{isDominated = \textbf{true}}{
					\textbf{continue}
				}
				$temp(w).add(l')$\\
				$queue.push(l')$
			}
		}
	}
	\BlankLine
	$paretoSet = \emptyset$\\
	\For{$\forall i \in I$}{
		$paretoSet(i) = perm(i)$
	}
	\Return{$paretoSet$}
	
\end{algorithm}

In order for the query phase to operate correctly on a cover graph, several adjustments must be made to standard MLS. The overlay graph is not necessarily simple, hence one transition between two neighboring cover nodes can result in different cost vectors (corresponding to different paths between the pair in the original graph). Additionally, the source vertex may not always belong to the cover. Besides that, t-discarding procedure should be optimized for the dimensionality reduction technique to decrease the query time as much as possible.

In cases when the source vertex is not a part of the kPC, it is connected to the overlay at the beginning of the query (line 5). This can be done similarly to the edge building procedure performed during the overlay construction. An enumerative DFS search operates from the source vertex on the original graph. Whenever it meets a cover vertex, the path is saved if it is not dominated by any other path.  If a goal vertex not included in the overlay is defined, it can be added to the overlay similarly. The only difference is that for incoming overlay edges to be found, the DFS search starting from the goal must traverse edges only backwards.

Due to the overlay graph having multiple cost vectors assigned to single edges, the modified search must iteratively expand not only to neighbor vertices, but also over single cost vectors which correspond to different compressed paths between two vertices (lines 14, 15). Therefore, while the overall number of iterations is decreased, the number of domination checks relative to the iteration number is significantly higher than that of MLS on the original graph. However, this circumstance further increases the relative speedup provided by dimensionality reduction, resulting in a synergy between the two techniques.

T-sets of the nodes are updated online: whenever a new label is added to the closed set, its truncated parameter vector is checked against the current t-set (lines 12, 13). If the candidate is not dominated by any of the included vectors, it is added to the t-set while simultaneously discarding all the vectors dominated by it. This approach preserves only mutually non-dominated truncated vectors in the t-sets, while minimizing the number of operations necessary. 

As can be seen, the query returns labels of path cost vectors belonging to the Pareto set. If every label contains a pointer to its parent, iterative steps through the pointers up to the source vertex are a simple way of extracting the compressed path corresponding to the desired solution. In order to be able to retrieve the full path, two minor adjustments must be made. Firstly, during the preprocessing phase, the algorithm must save not only the cost vectors of the overlay edges, but also the corresponding vertex sequences. Secondly, labels should be expanded to contain the ID of the edge cost vector that was used during the transition from the parent. Afterwards, for full path retrieval it suffices to add the intermediate nodes using the IDs before moving to the parent pointer.

\section{Evaluation}
The algorithm was implemented in C++ and compiled using GCC 9.3.0. The machine the experiments were run on operated on 4.2 GHz Intel i7-7700k with 32 GB RAM.
\subsection{Experiment setting}
The algorithm has been tested on two graph structures. The first graph represents the roadmap of Bavaria\protect\footnotemark \footnotetext{https://download.geofabrik.de/europe/germany/bayern.html}, Germany. In order to test the method on multiple criterion combinations of different sizes and correlation degrees, some of the criteria for the graph have been produced artificially. The time parameter $t$ provided in the source has been treated as the ground criterion used for the subsequent generation of the correlated criteria. While the structure of the graph has been left unaltered, the parameter vectors of edges have been replaced to create five datasets of varying nature. Table \ref{parameters} provides the overview of the criterion combinations that have been used for the experiments. The varying size of criterion vectors can be used to estimate the efficiency of t-discarding depending on whether the problem instance is bicriterial or not. Simultaneously, while the fully random criteria are not in any way correlated with the ground parameter, the formula $t + ct$ is expected to produce values somewhat close to those of the ground criterion. On the other hand, the inversely proportional combination $t$-$1/t$ represents a "heavy" bicriterial instance that is expected to produce Pareto sets of relatively large sizes.

The second graph used in the experiments is a representation of the San Francisco bay roadmap\protect\footnotemark \footnotetext{http://users.diag.uniroma1.it/challenge9/download.shtml}. The dataset is an unaltered instance of a real-world scenario considering two parameters: distance and traversal time. The sizes of both graphs are presented in Table \ref{sizes}.

For all datasets, the algorithm is set to minimize all of the objective criteria. Simultaneously, a criterion value of a path is calculated as the sum of corresponding values of its included edges for all criteria. The experiments have been repeated 1000 times on each of the datasets in order to obtain the average results. During every experiment, a source vertex and a subset of goal vertices have been picked at random from the whole dataset. The goal subsets are set to size 1000 on all datasets.

\begin{table}[t]
	\renewcommand{\arraystretch}{1.3}
	\caption{parameter combinations for the Bavaria graph. $t$ is the ground criterion. $r \in \{0, ..., 50\}$, $c \in [-0.5; 0.5]$ are randomly generated for every edge.}
	\begin{center}
		\label{parameters}
		\begin{tabular}{|c| l l l|}
			\hline
			name & $p_1$ & $p_2$ & $p_3$ \\
			\hline
			2-C & $t$ & $t + ct$ & - \\
			2-U & $t$ & $r$ & - \\
			3-C & $t$ & $t + ct$ & $t + ct$ \\
			3-U & $t$ & $r$ & $r$ \\
			$t$-$1/t$ & $t$ & $1/t$ & - \\
			\hline
		\end{tabular}
	\end{center}
\end{table}

\begin{table}[t]
	\renewcommand{\arraystretch}{1.3}
	\caption{graph sizes}
	\begin{center}
		\label{sizes}
		\begin{tabular}{|c| l l|}
			\hline
			graph & vertex \# & edge \#\\ 
			\hline
			Bavaria & 294727 & 587782 \\
			SF Bay & 321270 & 800172 \\
			\hline
		\end{tabular}
	\end{center}
\end{table}

\subsection{kPC evaluation}
First of all, the work of kPC technique has been tested. As the experiments described in \cite{kPC} show, setting the path size value $k$ above 32 generally results in only a minor performance gain while increasing the algorithm runtime dramatically. For this reason, the experiments have been conducted with $k=32$. The nodes have been processed in the increasing vertex ID order. Table \ref{kPC_sizes} shows the average overlay graph sizes. The vertex amount differed slightly in individual experiments due to randomly chosen goal vertices that must remain in the overlay graph. However, it is independent of the criteria maintained due to the fact that cover vertices are only selected based on the graph structure.

In contrast, the final number of overlay edges is influenced not only by the cover vertex layout, but also by the proportion of pruned cover edges. This proportion is in turn defined by the considered criteria. As expected for a fixed criterion number, the variety of cover edges is significantly decreased if the criteria are interconnected. On the other hand, the increase in criterion number results in a higher number of non-dominated cover edges, although, in this particular case, the correlation between the criteria has a stronger effect than their amount.

\begin{table}[t]
	\renewcommand{\arraystretch}{1.3}
	\caption{average sizes of overlay graphs for $k=32$}
	\begin{center}
		\label{kPC_sizes}
		\begin{tabular}{|c| l l l|}
			\hline
			dataset & vertex \# & edge \# & \begin{tabular}{@{}l@{}}construction\\time, s.\end{tabular}  \\
			\hline
			2-C & 23217 & 209394 & 62.08 \\
			2-U & 23234 & 268570 & 65.43 \\
			3-C & 23208 & 227797  & 62.49 \\
			3-U & 23225 & 313463  & 64.62 \\
			$t$-$1/t$ & 23213 & 313500 & 64.02 \\
			SF Bay & 35465 & 537250  & - \\
			\hline
		\end{tabular}
	\end{center}
\end{table}

\begin{figure*}
	\centering
	\begin{subfigure}[b]{0.49\linewidth}
		\includegraphics[width=\linewidth]{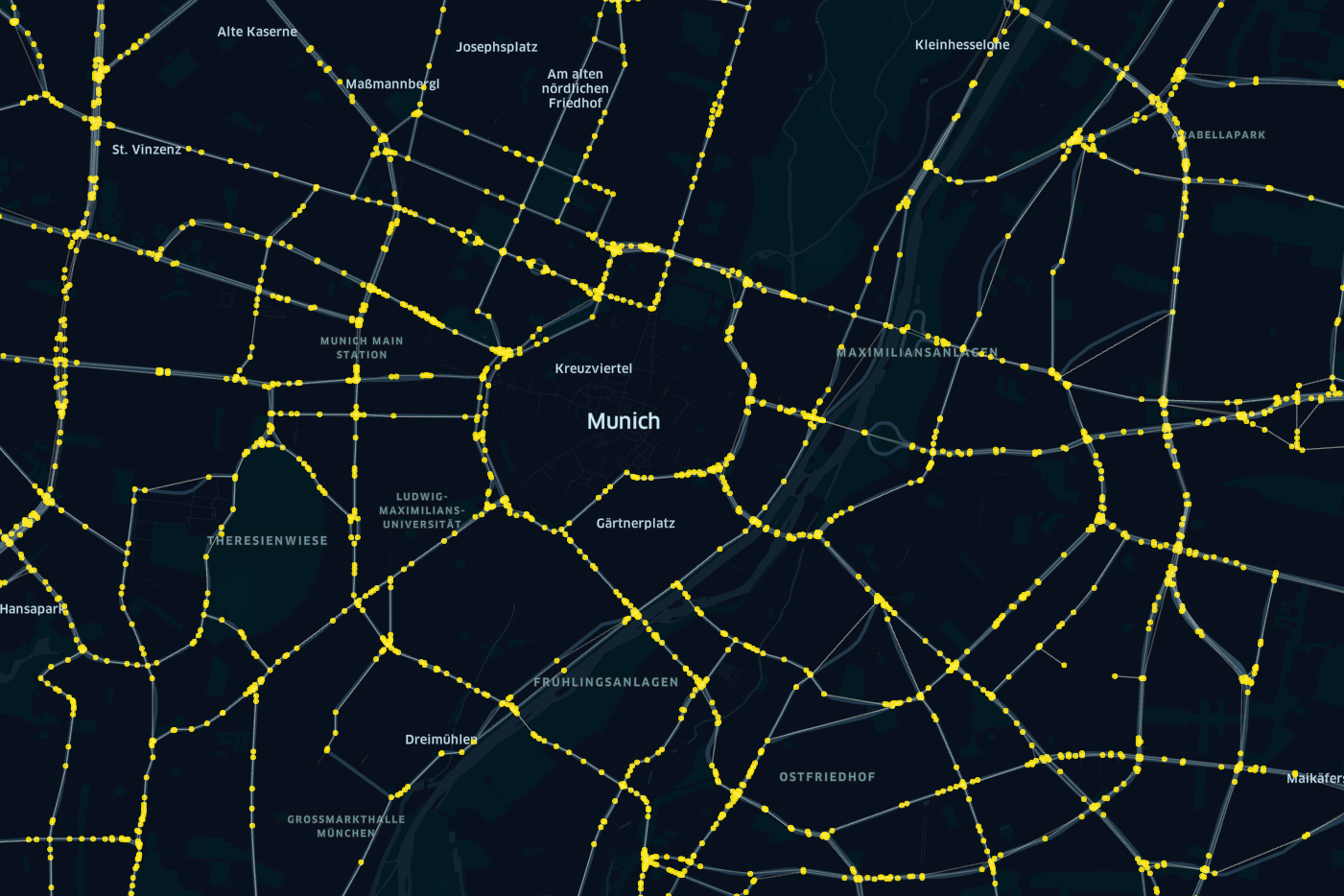}
		\caption{original graph}
	\end{subfigure}
	\begin{subfigure}[b]{0.49\linewidth}
		\includegraphics[width=\linewidth]{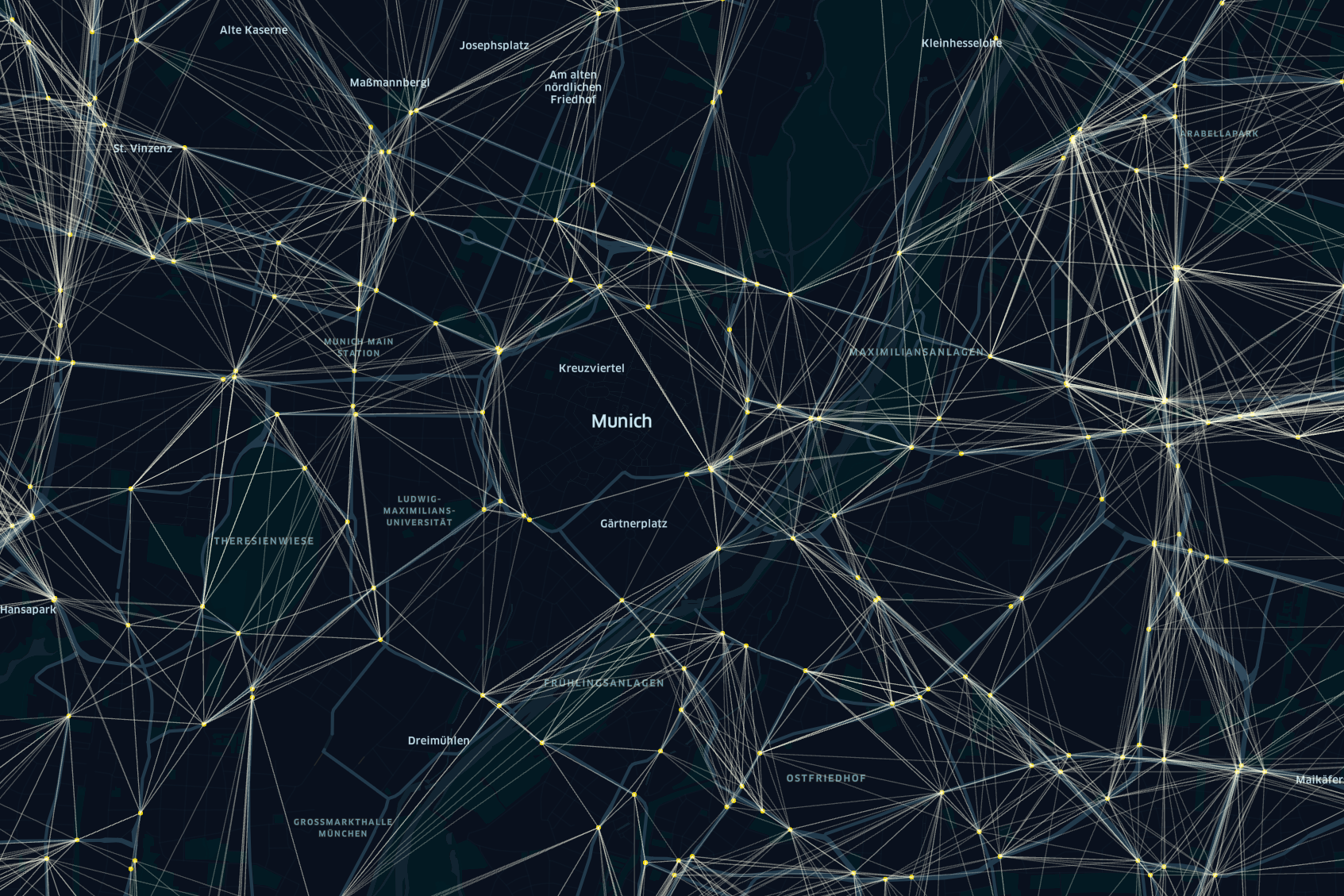}
		\caption{$k$PC graph for $k=32$}
	\end{subfigure}
	\caption{Bavaria graph visualization parts produced using \textit{kepler.gl}. Yellow dots represent vertices, white lines are graph edges.}
	\label{fig:graph_pics}
\end{figure*}

It can be seen from the table that the vertex quantity in the covers for all of the datasets is close to $10\%$ of the original vertex number, and the number of cover edges comprises only a fraction of the original ones. On the other hand, while the cover graph construction for every Bavaria dataset is performed in approximately 1 minute, the process for the SF dataset can last anywhere from 20 to 200 hours, varying significantly in duration based on the goal vertex set. The main reason for this is the higher density of SF graph, which results in the exponential growth of generated cover edges. Nevertheless, the majority of these edges are dominated and therefore pruned by the pruning techniques. Fortunately, these increased time requirements are of no significant importance since the preprocessing only has to be performed again if the combinatorial structure of the graph is changed. However, if the cover graph must be built in a shorter time, one can decrease the $k$ value of the algorithm. For $k=24$, the cover graph of SF bay is built in under 20 minutes and consists of 40k vertices and 520k edges on average, providing a graph of approximately the same size at only a fraction of time. In fact, this option appears to be more preferable, since the processing time and edge number are decreased dramatically while only increasing the amount of vertices by $15\%$.

\subsection{Optimal runtime evaluation}
\begin{table*}
	\renewcommand{\arraystretch}{1.3}
	\caption{average optimal solution runtimes and their standard deviations in seconds}
	\begin{center}
		\label{runtimes}
		\begin{tabular}{|c|l l l l|}
			\hline
			dataset & MLS & $t$-MLS & kPC-MLS &$t$-kPC-MLS \\
			\hline
			2-C & $4.97 \pm 4.68$ & $5.97 \pm 4.50$ & $\mathbf{0.83 \pm 0.62}$ & $1.20 \pm 0.79$ \\
			2-U & $1778.69 \pm 1644.70$ & $319.65 \pm 183.31$ & $696.17 \pm 713.60$ & $\mathbf{66.59 \pm 40.28}$ \\
			3-C & $76.36 \pm 190.39$ & $78.12 \pm 157.09$ & $\mathbf{7.86 \pm 17.00}$ & $10.21 \pm 20.70$ \\
			3-U & - & - & $302777.64 \pm 186906.40$ & $\mathbf{10362.24 \pm 7911.91}$ \\
			$t$-$1/t$ & $16421.45 \pm 5736.12$ & $1534.00 \pm 1118.89$ & $7672.59 \pm 2859.86$ & $\mathbf{270.77 \pm 54.66}$ \\
			SF Bay & $2194.38 \pm 1369.69$ & $357.51 \pm 141.56$ & $1179.86 \pm 865.38$ & $\mathbf{107.63 \pm 44.47}$ \\
			\hline
		\end{tabular}
	\end{center}
\end{table*}

\begin{table}
	\renewcommand{\arraystretch}{1.3}
	\caption{ratios of accelerations provided by individual techniques, computed as $T/t$, where $T$ is the average operation time of vanilla MLS on the dataset and $t$ is the time of the corresponding technique.}
	\begin{center}
		\label{runtime_coeffs}
		\begin{tabular}{|c|l l l l|}
			\hline
			dataset & MLS & $t$-MLS & kPC-MLS &$t$-kPC-MLS \\
			\hline
			2-C & 1.00 & 0.83 & \textbf{5.99} & 4.14 \\
			2-U & 1.00 & 5.56 & 2.55 & \textbf{26.71} \\
			3-C & 1.00 & 0.98 & \textbf{9.71} & 7.48 \\
			3-U & - & - & 1.00 & \textbf{29.22} \\
			$t$-$1/t$ & 1.00 & 10.71 & 2.14 & \textbf{60.64} \\
			SF Bay & 1.00 & 6.13 & 1.86 & \textbf{20.39} \\
			\hline
		\end{tabular}
	\end{center}
\end{table}

In order to properly analyze the influence of individual parts of the combined algorithm, each of the 4 possible combinations has been tested on all of the datasets. To make the experiments fair, every experiment started with a randomly chosen source vertex, from which every technique combination has been run one after another. For all datasets except the SF bay graph, the corresponding covers have been generated using $k=32$. As for the kPC version of SF bay, it has been produced by $k=24$ to measure the algorithm performance under limited preparation time. The average runtimes and respective standard deviations on all the datasets are provided in the table \ref{runtimes}.

As expected, classic MLS is the slowest of the method combinations considered in the research. Therefore, it has been used as a benchmark for other tested techniques. Unfortunately, it was not possible to properly analyze the efficiency of vanilla MLS and $t$-MLS on the 3-U dataset using the available hardware due to insufficient RAM volumes. However, this is not the case for kPC-using approaches. The biggest part of memory used during path queries is used to store the permanent and temporary label sets. Due to the cover graphs produced through kPC consisting of only a fraction of the original vertex sets (approx. $10\%$), the amount of memory used decreases accordingly. Another notable circumstance visible from the table are rather high variance values across all datasets. This is caused by a significant presence of outliers, some of which exceed the average runtime values by tens of times. Result examination shows that these outliers are produced by queries starting from far corners of the graphs.

As Tables \ref{runtimes} and \ref{runtime_coeffs} show, despite being extremely efficient under certain conditions, $t$-discarding is not always of use. On instances with heavily correlated criteria, it somewhat slows queries down. This can be explained by two circumstances. Firstly, the Pareto set sizes for correlated criteria are relatively small, therefore the search algorithm does not spend as much time performing domination checks as it does for uncorrelated criteria. This makes the acceleration provided by $t$-discarding negligible. Secondly, maintenance and regular updates of truncated vector sets require the algorithm to perform additional operations, which causes $t$-discarding to perform worse than standard domination verification. However, this is only the case for heavily correlated parameters, present in datasets 2-C and 3-C. On all other datasets $t$-discarding demonstrates notable effectiveness, accelerating the queries by at least 5 times. On $t$-$1/t$, the heaviest two-criteria dataset of the tested ones, the employment of $t$-discarding alone reduces query runtimes by over 10 times. The conclusion drawn from these observations is while $t$-discarding can be of great help, it should be applied carefully when working with heavily correlated criteria.

On the other hand, kPC provides more stable speedup results, though not as high as $t$-discarding. The minimum speedup provided by kPC has been achieved on the SF bay dataset. This can be explained by the fact that the number of cover vertices and edges relative to the original one is higher than for the Bavaria graph. Interestingly, kPC provides a significantly higher speed gain on correlated datasets, which can be explained by the fact that the corresponding graphs have lower relative number of edges. Nevertheless, the experiments show that kPC influences the query performance positively on all datasets, although the nature of the method suggests it would be less helpful on small graphs. An important quality is that kPC significantly decreases not only time requirements of the searching algorithm, but also the amount of necessary memory. Since the memory is mostly used to store the path labels, a query operating on a cover graph containing only a small part of the original vertices would have several times lower memory requirements. In fact, a cover graph consisting of only $10\%$ of the original vertices would yield on average 10 times less labels. In practice, the expected memory saving of kPC-MLS compared to vanilla MLS would be somewhat lower due to the necessity of storing the cover graph in addition to the original one.

Being combined, $t$-discarding and kPC amplify their individual results. The tendency of $t$-discarding being inefficient on heavily correlated instances persists, but this result is mitigated by kPC, and the combination of these techniques is still dramatically more efficient than class MLS. As one can expect, the highest gain is achieved if both techniques are able to provide some acceleration individually. Combined, these form a synergy that results in far better results than the sum of their individual gains. The lowest gained acceleration being achieved on SF bay dataset exceeds 20 times over vanilla MLS. For the "heavy" instance $t$-$1/t$, the combination accelerates queries by 60 times while still maintaining solution optimality. Although there is no possibility to compare $t$-kPC-MLS to standard MLS on 3-U dataset, the difference between kPC-MLS and $t$-kPC-MLS in this case being similar to that on $t$-$1/t$ leads to an assumption of the gain being close to 60 or higher. These results indicate that the techniques indeed operate in a synergy that can be briefly described as follows: kPC decreases the number of vertices and labels to be processed, while $t$-discarding accelerates domination checks and mitigates the deceleration caused by their number being increased by kPC.

\subsection{Memory requirements evaluation}
Due to the sizes of Pareto sets having exponential dependency on graph sizes, the amount of memory required for queries on large graphs should also be evaluated and carefully considered. The two main structures managed in memory during multicriteria shortest path queries are the graph itself and Pareto sets of vertices maintained in the temporary and permanent sets. If a kPC cover of the graph is built, the memory required for its storage should also be considered, although it is approximately equal to that of the original graph. However, these amounts of memory remain constant for the graphs, while the sizes of Pareto sets during queries and their respective memory requirements may vary significantly. Therefore, average sizes of Pareto sets for the tested instances have also been measured.

\begin{table}
	\renewcommand{\arraystretch}{1.3}
	\caption{sizes of Pareto sets in millions on graphs and their covers, presented in form \textit{average/maximum}}
	\begin{center}
		\label{memory}
		\begin{tabular}{|c|l l|}
			\hline
			dataset & original & kPC \\
			\hline
			2-C & 2.283/10.497 & 0.182/0.835 \\
			2-U & 52.689/154.288 & 4.157/11.888 \\
			3-C & 12.145/81.538 & 0.972/6.381 \\
			3-U & - & 86.834/ 204.304 \\
			$t$-$1/t$ & 190.240/387.660 & 16.387/28.485 \\
			SF Bay & 62.681/186.510  & 7.894/31.830 \\
			\hline
		\end{tabular}
	\end{center}
\end{table}

Table \ref{memory} provides the estimates of average and maximum Pareto set sizes produced during the experiments. It must be noted that $t$-discarding alters neither the number of processed vertices nor the actual Pareto sets. However, it may require additional memory for operation depending on the way it is implemented. Nevertheless, even if its truncated vectors are stored permanently, the amount of memory they occupy is smaller than that of Pareto sets due to their decreased size and numbers.

As the table shows, apart from providing substantial speedup, kPC decreases memory requirements of the queries multiple times. As one can expect, the reduction in label number roughly corresponds to the ratio of vertices that remain in the kPC cover. This decrease in memory requirements creates another type of synergy between kPC and $t$-discarding, compensating for the memory consumed by truncated vectors. However, it is important to note that time and memory requirements of singular queries can vary significantly, which is shown by the high variance of optimal runtimes and the maximum label numbers being several times higher than their average counterparts. At the moment of writing the article, no reliable way has been found to predict the degree of variance except for the empirical one. Therefore, a series of test runs must be conducted on a new graph to estimate the maximum amount of memory the search algorithm may require.

\section{Conclusion}
The paper introduces a novel combination of techniques for accelerated one-to-many multicriteria shortest path search. During the experiments, the approach has been compared to a classic multicriteria search algorithm on multiple datasets containing criteria combinations of varied sizes and correlation degrees. Providing a reliable basis for the conclusions on its performance, the tests show it gives at least 6 times speedup on simple problem instances. Moreover, the time gain increases with the complexity of the dataset and can even exceed 60 times while still returning optimal Pareto sets. Exclusion of irrelevant vertices from queries through kPC preprocessing decreases memory requirements on average by 13 times. Due to the core query algorithm being a modified version of MLS, the approach can additionally adopt existing heuristics for further acceleration.

\bibliographystyle{asmems4}

\bibliography{preprint_version}

\end{document}